# HIGH-QUALITY IMAGE INTERPOLATION VIA LOCAL AUTOREGRESSIVE AND NONLOCAL 3-D SPARSE REGULARIZATION

*Xinwei Gao[1], Jian Zhang[1], Feng Jiang[1], Xiaopeng Fan[1], Siwei Ma[2], Debin Zhao[1]*

[1]School of Computer Science and Technology, Harbin Institute of Technology, Harbin, China
[2]School of Electronic Engineering & Computer Science, Peking University, Beijing, China
Email：{xwgao.cs, jzhangcs, fjiang, fxp, dbzhao}@hit.edu.cn, {swma}@pku.edu.cn

**ABSTRACT**

In this paper, we propose a novel image interpolation algorithm, which is formulated via combining both the local autoregressive (AR) model and the nonlocal adaptive 3-D sparse model as regularized constraints under the regularization framework. Estimating the high-resolution image by the local AR regularization is different from these conventional AR models, which weighted calculates the interpolation coefficients without considering the rough structural similarity between the low-resolution (LR) and high-resolution (HR) images. Then the nonlocal adaptive 3-D sparse model is formulated to regularize the interpolated HR image, which provides a way to modify these pixels with the problem of numerical stability caused by AR model. In addition, a new Split-Bregman based iterative algorithm is developed to solve the above optimization problem iteratively. Experiment results demonstrate that the proposed algorithm achieves significant performance improvements over the traditional algorithms in terms of both objective quality and visual perception.

## 1. INTRODUCTION

Image interpolation, which aims to recover a high-resolution (HR) image $x$ from its downsampled low-resolution (LR) version $y$, has become a very active research area in image processing [1-10] with a wide range of applications in digital photography, video communication, medical analysis, object recognition, satellite remote sensing, and consumer electronics. For an observed image y, image interpolation can be expressed by

$$y = Dx + n, \quad (1)$$

where x is the high-resolution image to be estimated, D is a downsampling operator, and $n$ is the additive noise vector. It can be started out by observing that image interpolation is a typical ill-posed inverse problem. So the solution to Eq. (1) with the $l_2$ norm fidelity constraint is general and direct:

$$\tilde{x} = \underset{x}{argmin} \|y - Dx\|_2^2. \quad (2)$$

As a classic method, the autoregressive (AR) model has performed well in image interpolation [2-4], which is locally computed from a low-resolution image by the least square model and well reconstruct the high-resolution edge structures:

$$x_i = \sum_{k=1}^{n} w_s^k x_i^{(k)} + e_i, \quad (3)$$

where w is the regressive coefficients vector, $x_i$ is the gray level of one pixel, and $x_i^{(k)}$ is the *k*-th corresponding neighbor in an given lexicographical order. When the regressive coefficients vectors for these pixels in a local window *S* are assumed to be the same, we write Eq. (3) in matrix form:

$$x_S = A_S w_S + E_S, \quad (4)$$

where $X_S = [x_1, x_2, \cdots, x_N]^T$ is a column vector containing all these *N* pixels inside the local square window *S*. $w_S = [w_S^1, w_S^2, \cdots, w_S^n]^T$ is the vector of the AR coefficients. The *j*-th row of the matrix $A_S$ are the lexicographical ordered neighbors of the pixel $x_j$. The linear system x = Aw having close form solution can be solved by least squaring this problem:

$$w_S = A_S^+ x_S, \quad (5)$$

where $A^+ = (A^T A)^{-1} A^T$ stands for the Moore-Penrose pseudoinverse of matrix A. Then the structural similarity between LR and HR images is used for estimating the HR image $\tilde{x}$. One important issue regarding the AR model is numerical stability, which is related to the rank condition of the design matrix. On one hand, if the AR model order $n$ (i.e., the length of the AR coefficients vector) is large, which means there are many parameters we should estimate, and when the number of the equations for these variables is not enough, the problem of numerical stability will emerge. On the other hand, if pixels in the local AR window have weak spatial correlation with neighboring pixels, such as the window is an occlusion region, the accuracy of estimation will decline a large due to the interposition of uncorrelated spatial neighbors. On a second reflection, fortunately, the two dimensions of the image signal offer ways to bypass the problem of numerical stability. One way to increase the number of equations on pixels in the window is to iteratively use both the LR samples and the estimated HR ones. Anoth-er way is to introduce regularized constraints on frequency domains on the image interpolation.

According to the two ways above, we propose a novel image interpolation algorithm, which is formulated via combining both the local autoregressive (AR) model and the nonlocal adaptive 3-D sparse model as regularized constraints under the regularization framework. Estimating the high-resolution image by the local AR regularization is different from these conventional AR models, which weighted calculates the interpolation coefficients without considering the rough structural similarity between the low-resolution (LR) and high-resolution (HR) images. Then the nonlocal adaptive 3-D sparse model is formulated to regularize the interpolated HR image, which provides a way to modify these pixels with the problem of numerical stability caused by AR model. In addition, a new Split-Bregman based iterative algorithm is developed to solve the above optimization problem iteratively. Experiment results demonstrate that the proposed algorithm achieves significant performance improvements over the traditional algorithms in terms of both objective quality and visual perception.

The rest of this paper is organized as follows. Section 2 presents the proposed algorithm. Implementation details and experimental results are shown in Section 3 and Section 4. Section 5 concludes this paper.

## 2. THE PROPOSED ALGORITHM UTILIZING LOCAL AND NONLOCAL REGULARIZATION

In this section, we first introduce the local iterative AR model. Then the nonlocal adaptive 3-D sparse representation is presented. Finally, the effective solution of the optimization is proposed.

### 2.1. Local AR regularized constraint

In recent years, autoregressive model is one of the most popular and widely used models for the image interpolation, which is locally computed from a LR image by the least square model and well reconstructs the edge of HR image. To model the structural similarity between the LR and HR images, the pixel in the LR image is assumed to share the same AR coefficients with the corresponding pixel in the HR image:

$$w_S = \underset{w}{\operatorname{argmin}} \left\| y_{S_L} - A_{S_L} w \right\|_2^2, \quad (6)$$

$$\tilde{x}_{S_H} = A_{S_H} w_S, \quad (7)$$

where $w_S$ is the AR coefficients vector, which is calculated on the LR image and is used to estimate the HR image. However, the similarity of geometric duality between the LR image and the corresponding HR image is not accurate occasionally. Different from the conventional AR model, we propose a weighted AR model as the local regularization, which estimates both the AR coefficients vector and the interpolated HR image only on the initialized HR image. In addition we enhance the AR model's numerical stability by increasing the number of equations on pixels in the larger coefficient training window $S$. The proposed local AR regularization $\Phi(x,w)$ is shown:

$$\Phi(x,w) = \sum_{i \in S} \left\{ \sum_{k=1}^{n} \theta(i,k) \left| x_i^{(k)} - \sum_{j=1}^{n} w_j x_{i,k}^{(j)} \right|^2 \right\}, \quad (8)$$

where $\theta(i,k)$ means the weight to describe the similarity texture between the patches with center of the pixel $x_i$ and its neighbor $x_i^{(k)}$.

$$\theta(i,k) = \frac{1}{Z(i)} e^{-\mu \left\| (P_i - P_i^{(k)}) \cdot L(i,k) \right\|_2^2}, \quad (9)$$

$$Z(i) = \sum_{i \in S} e^{-\mu \left\| (P_i - P_i^{(k)}) \cdot L(i,k) \right\|_2^2}, \quad (10)$$

In Eqs. (9) and (10), $P_i$ and $p_i^{(k)}$ are the matrices whose elements are pixel values in the patches with center of $x_i$ the $k$-th corresponding neighbor $x_i^{(k)}$ in HR image, $L(i,k)$ expresses the squared Euclidean distance, and $\mu$ is a constant to control the decay of the exponential function.

### 2.2. Nonlocal 3-D sparse regularized constraint

Natural images are not random 2-D signals, which are only covered by some sparse flow patterns of the 2-D signal space, and the method of nonlocal means [11] is successfully used in image processing. We introduce a nonlocal adaptive 3-D sparse model to better characterize the nonlocal image model as shown in Fig. 1. These overlapped blocks, which are similar with the block $b_q$ centered on the current interpolated pixel $\tilde{x}_q$, are searched and put into a block set $S_q$,

$$S_q = \left\{ b_j = b_1, b_2, \cdots, b_n \; s.t. \; \left\| b_q - b_j \right\|_2^2 < \varepsilon \right\}, \quad (11)$$

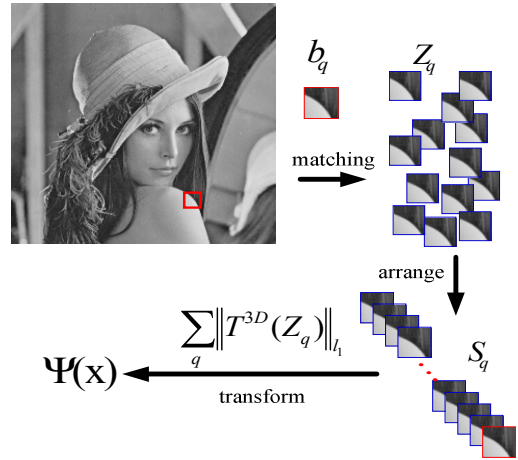

**Fig. 1**. Nonlocal adaptive 3-D sparse model

where $\varepsilon$ is the similarity threshold. Then, these blocks are arranged into a 3-D sequence, defined by $Z_q$. The function of the nonlocal adaptive 3-D sparse interpolation model is written as

$$\Psi(x) = \sum_q \left\| T^{3D}(Z_q) \right\|_1. \quad (12)$$

In Eq. (12), $T^{3D}$ is a 3-D transform operator. Different from BM3D [12] and [13], the nonlocal sparse interpolation method, combining the sparse interpolation and the nonlocal means, is based on the HR estimated image created by local iterative AR model.

### 2.3. Image interpolation via collaborative local and nonlocal regularization method

Incorporating (7) and (11) into $\tilde{x} = \underset{x}{argmin} \|y - Dx\|_2^2$, the whole framework of the proposed method is interpreted as a combination of three parts: data fidelity term, a local regularization term and a nonlocal regularization term:

$$(x, w) = \underset{x,w}{argmin} \|y - Dx\|_2^2 + \lambda \cdot \Phi(x, w) + \gamma \cdot \Psi(x), (13)$$

where $\lambda$ and $\gamma$ are control parameters. Therefore, it is our belief that better results will be achieved by imposing the above three constraints into the ill-posed problem of image interpolation. However, the above optimization problem is essentially quite difficult to solve directly due to the non-differentiability and non-linearity of the nonlocal term. In the next section, the implementation details of this image interpolation method are provided.

### 3. IMPLEMENTATION DETAILS

In recent years, a new iteration algorithm [15] called Split-Bregman efficiently solves part of the $l_1$ formed minimization, and in the proposed method, a modified iterative algorithm for solving Eq. (13) is developed. Instead of solving the optimization problem directly, Eq. (13) can be first transformed into an equivalent variant by introducing auxiliary variables g and h:

$$\begin{cases} \underset{x,g,h,w}{argmin} \frac{1}{2} \|y - Dx\|_2^2 + \lambda \cdot \Phi(g, w) + \gamma \cdot \Psi(h) \\ s.t. \quad x = g, \quad x = h \end{cases} . \quad (14)$$

Applying this iteration algorithm of Split-Bregman, The corresponding functions of (14) can be broken up into the following three iterative steps:

$$(\tilde{x}^{(t)}, \tilde{g}^{(t)}, \tilde{h}^{(t)}, \tilde{w}^{(t)}) = \underset{x,g,h,w}{argmin} \frac{1}{2} \|y - Dx\|_2^2 + \lambda \cdot \Phi(g, w)$$
$$+ \gamma \cdot \Psi(h) + \alpha \|x - g - U^{(t-1)}\|_2^2 + \beta \|x - h - V^{(t-1)}\|_2^2 \quad (15)$$

$$U^{(t)} = U^{(t-1)} - (\tilde{x}^{(t)} - \tilde{g}^{(t)}), \quad (16)$$

$$V^{(t)} = V^{(t-1)} - (\tilde{x}^{(t)} - \tilde{h}^{(t)}), \quad (17)$$

where $\alpha$ and $\beta$ are the parameters associated with the constraints x = g and x = h, $t$ means the $t$-th iteration. By observing Eq. (15), it can be performed efficiently by the alternating minimization with respect to x, w, g and h separately. In the following, we present an efficient solution for every separated sub-problem.

### 3.1. x sub-problem

Given $\tilde{w}^{(t-1)}$, $\tilde{g}^{(t-1)}$ and $\tilde{h}^{(t-1)}$, the optimization problem associated with $x$ can be expressed as

$$\tilde{x}^{(t)} = \underset{x}{argmin} \frac{1}{2} \|y - Dx\|_2^2 + \alpha \|x - \tilde{g}^{(t-1)} - U^{(t-1)}\|_2^2$$
$$+ \beta \|x - \tilde{h}^{(t-1)} - V^{(t-1)}\|_2^2 , \quad (18)$$

where each term of the Eq. (18) is quadratic term. So it is a minimization problem of strictly convex quadratic function, which closed form solution for $\tilde{x}^{(t)}$ can be expressed as:

$$\tilde{x}^{(t)} = \left(D^T D + (\alpha + \beta) I\right)^{-1} \cdot r^{(t-1)}, \quad (19)$$

where $r^{(t-1)} = D^T y + \alpha(\tilde{g}^{(t-1)} + U^{(t-1)}) + \beta(\tilde{h}^{(t-1)} + V^{(t-1)})$, I is identity matrix.

### 3.2. w and g sub-problem

To minimize $\Phi(g, w)$, we iteratively update the AR coefficients vector and the interpolated HR pixels in turn:

$$\tilde{w}^{(t)} = \underset{w}{argmin} \sum_{i \in S} \left\{ \sum_{k=1}^{n} \theta(i,k) \left| g_i^{(k)} - \sum_{j=1}^{n} w_j g_{i,k}^{(j)} \right|^2 \right\}, \quad (20)$$

$$\tilde{g}^{(t)} = A^{(t-1)} w^{(t)}, \quad (21)$$

where $\tilde{w}^{(t)}$ is solved by weighted least squaring this sub-problem, $\theta(i,k)$ is defined by Eq. (9), and then $\tilde{g}^{(t)}$ is generated as the linear weighted summation with the regressive coefficients vector $\tilde{w}^{(t)}$.

### 3.3. h sub-problem

Given $\tilde{y}^{(t-1)}$, $\tilde{g}^{(t-1)}$, and $\tilde{w}^{(t-1)}$, the optimization problem associated with $\tilde{h}^{(t)}$ can be writen as

$$\tilde{h}^{(t)} = \underset{h}{argmin} \left\{ \Psi(h) + \beta \|\tilde{y}^{(t-1)} - h - V^{(t-1)}\|_2^2 \right\}. \quad (22)$$

The proximal map associated to the large number of functions can only be solved with an approximation. This proximal map associated to $\Psi(h)$ in Eq. (22) is calculated by applying the algorithm of shrinkage in transform domain of all the 3-D arrays centered at each pixel in the HR image.

### 3.4. Summary of the implementation

So far, the whole framework in the process of handing the sub-problems has been solved. In light of all the derivations above, Table 1 outlines the description of the proposed image interpolation algorithm.

**Table 1.** Description of the proposed algorithm

**Input:** the LR image y, and $\lambda, \gamma, \alpha, \beta$.
**Initialization:**
Set $t=0$, $U^{(0)} = V^{(0)} = 0$, $\tilde{g}^{(0)} = \tilde{h}^{(0)} = 0$.
Estimate HR image $\tilde{x}^{(0)}$ by Bicubic interpolation filter;
**Loop:**
Iterate on $t=1$ to $t = Max\_iterative$:

$$\tilde{x}^{(t)} = \left(D^T D + (\alpha+\beta)I\right)^{-1} \cdot r^{(t-1)} \quad (19)$$

$$\tilde{w}^{(t)} = \underset{w}{argmin} \sum_{i \in S}\left\{\sum_{k=1}^{n} \theta(i,k)\left|g_i^{(k)} - \sum_{j=1}^{n} w_j g_{i,k}^{(j)}\right|^2\right\} \quad (20)$$

$$\tilde{g}^{(t)} = A^{(t-1)} w^{(t)} \quad (21)$$

$$\tilde{h}^{(t)} = \underset{h}{argmin}\left\{\Psi(h) + \beta\left\|\tilde{y}^{(t-1)} - h - V^{(t-1)}\right\|_2^2\right\} \quad (22)$$

$$U^{(t)} = U^{(t-1)} - (\tilde{x}^{(t)} - \tilde{g}^{(t)}) \quad (17)$$

$$V^{(t)} = V^{(t-1)} - (\tilde{x}^{(t)} - \tilde{h}^{(t)}) \quad (18)$$

**Output:** the HR interpolated image $\tilde{x}$.

## 4. EXPERIMENTAL RESULTS

To evaluate the efficacy of the proposed algorithm, extensive experiments were carried out in this section. For thoroughness and fairness of our comparison study, we exploit some widely used eight test images: *Airplane*, *Boat*, *Cap*, *Door*, *Girl*, *Lena*, *Monarch*, and *Peppers* of the size 512x512 as shown in Fig. 2. Following the conventional setting, we first downsample these HR images to get the LR images by a scale factor of 2 in both row and column coordinate axis. The estimated HR images are got by the proposed method and the comparative methods. The performance is measured by PSNR and FSIM [14] between the original image and the corresponding interpolated image. Our method is compared with some representative work in the literature: (1) Bicubic interpolation [1], (2) new edge-directed interpolation (NEDI) [2], (3) nonlocal edge-directed interpolation (NLEDI) [3], (4) SAI interpolation [4], and (5) Robust SAI [16], and the results are all generated by the codes of original authors with the respective optimized parameters. The operator of 3-D transform $T^{3D}$ iscombined by a 2-D discrete wavelet transform (DWT) and a 1-D discrete cosine transform (DCT).

It can be observed in Table 2 that the proposed algorithm performs better than the other methods for all these test images consistently. Compared with a global method Bicubic, the proposed method can improve the objective quality of the generated images. The average gain is more than 1dB. Our method outperforms the edge detection based local method NEDI, for which the average gains is 0.90dB. Compared with a nonlocal method of NLEDI, the average gain is 1.20dB. Our method is more effective than SAI and Robust SAI, which are based on the autoregressive model. The gains of the proposed method are 0.45dB and 0.31dB compared with SAI and Robust SAI respectively.

The PSNR index can measure the intensity difference between two images, but it may fail to describe the visual perception quality. Some IQAs are used measures for image visual quality assessment. Recently, a new image quality assessment (IQA) model FSIM is proposed, which *achieves much higher consistency with the subjective evaluations than state-of-the-art IQA metrics* [14]. The higher FSIM value means the better visual quality. From Tables 2, it could be seen that proposed algorithm again achieves the highest average FSIM scores among the competing methods. It means that our method can achieve better performance on the image visual quality. It can be seen that the proposed method produces better visually pleasant results among these competing methods in Fig. 3, Fig. 4 and Fig. 5.

## 5. CONCLUSIONS

This paper presents a novel image interpolation algorithm, which is formulated via combining both the local autoregressive (AR) model and the nonlocal adaptive 3-D sparse model as regularized constraints under the regularization framework. Estimating the high-resolution image by the local AR regularization is different from these conventional AR models, which weighted calculates the interpolation coefficients without considering the rough structural similarity between the low-resolution (LR) and high-resolution (HR) images. Then the nonlocal adaptive 3-D sparse model is formulated to regularize the interpolated HR image, which provides a way to modify these pixels with the problem of numerical stability caused by AR model. In addition, a new Split-Bregman based iterative algorithm is developed to solve the above optimization problem iteratively. Experiment results demonstrate that the proposed algorithm achieves significant performance improvements over the traditional algorithms in terms of PSNR, FSIM and visual quality.

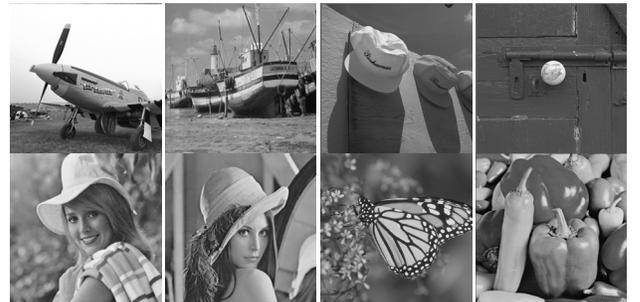

**Fig. 2**. The testing image set.

**Table 2.** The PSNR and FSIM comparison results of these methods

| Image | Bicubic | | NEDI [2] | | NLEDI [3] | | SAI [4] | | Robust SAI [16] | | Proposed | |
|---|---|---|---|---|---|---|---|---|---|---|---|---|
| | PSNR | FSIM | PSNR | FSIM | PSNR | FSIM | PSNR | FSIM | PSNR | FSIM | PSNR | FSIM |
| *Airplane* | 29.61 | 0.9787 | 29.97 | 0.9787 | 29.52 | 0.978 | 30.62 | 0.9817 | 30.64 | 0.9819 | **30.84** | **0.9823** |
| *Boat* | 29.21 | 0.9695 | 29.29 | 0.9675 | 29.10 | 0.9687 | 29.66 | 0.9712 | 29.77 | 0.9719 | **29.89** | **0.9729** |
| *Cap* | 32.21 | 0.9753 | 32.55 | 0.9750 | 32.18 | 0.9744 | 33.00 | 0.9773 | 33.14 | 0.9778 | **33.32** | **0.9779** |
| *Door* | 31.75 | 0.9584 | 31.66 | 0.9571 | 31.89 | 0.9596 | 31.88 | 0.9591 | 32.06 | 0.9603 | **32.13** | **0.9605** |
| *Girl* | 31.48 | 0.9627 | 32.05 | 0.9704 | 31.75 | 0.9679 | 31.71 | 0.9642 | 32.02 | 0.9680 | **32.77** | **0.9681** |
| *Lena* | 33.91 | 0.9874 | 33.77 | 0.9869 | 33.35 | 0.9857 | 34.67 | 0.9884 | 34.72 | 0.9889 | **34.95** | **0.9893** |
| *Monarch* | 29.28 | 0.9713 | 29.34 | 0.9716 | 28.85 | 0.9709 | 30.28 | 0.9747 | 30.39 | 0.9761 | **30.99** | **0.9803** |
| *Peppers* | 32.79 | 0.9785 | 33.08 | 0.9809 | 32.65 | 0.9793 | 33.53 | 0.9806 | 33.69 | 0.9819 | **34.01** | **0.9838** |
| Average | 31.28 | 0.9727 | 31.46 | 0.9735 | 31.16 | 0.9730 | 31.91 | 0.9746 | 32.05 | 0.9758 | **32.36** | **0.9764** |

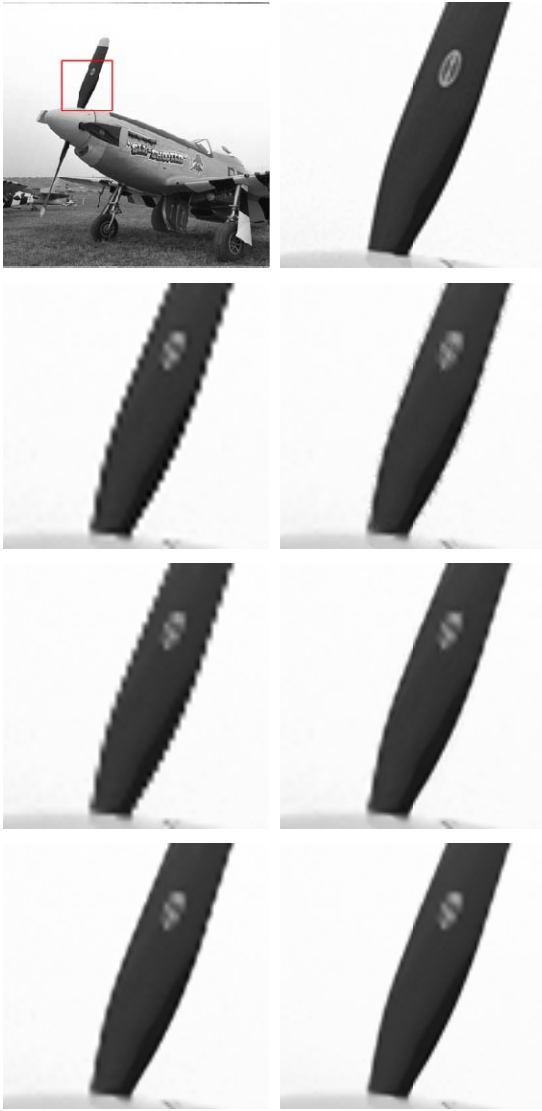

**Fig. 3**. The visual quality comparison on *Airplane*. Methods from left to right and then from top to bottom are original image, original block, and the blocks interpolated by Bicubic, NEDI, NLEDI, SAI, Robust SAI and the proposed algorithm respectively. The interpolation results of the proposed algorithm present the highest qualities.

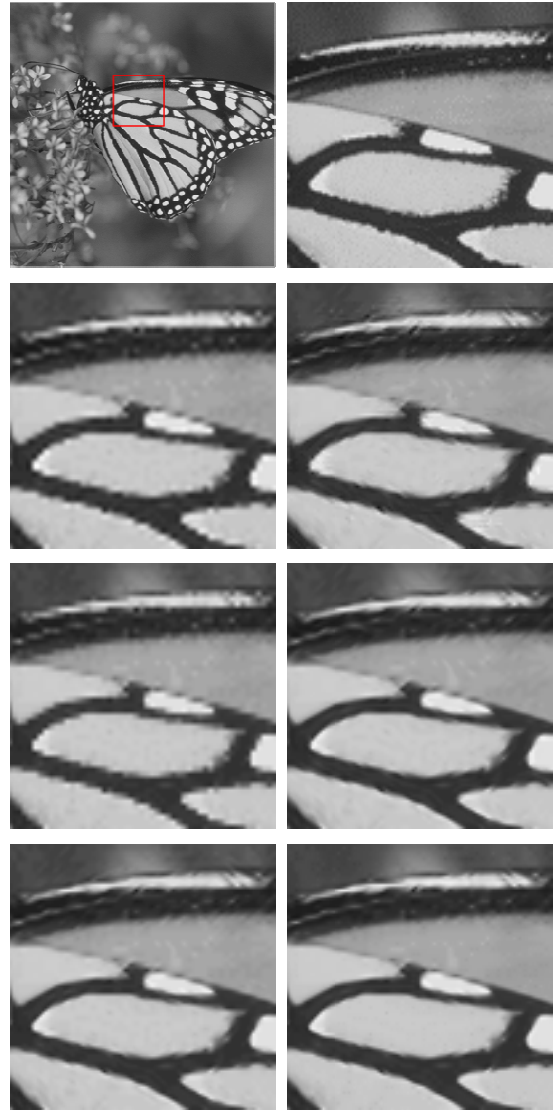

**Fig. 4**. The visual quality comparison on *Peppers*. Methods from left to right and then from top to bottom are original image, original block, and the blocks interpolated by Bicubic, NEDI, NLEDI, SAI, Robust SAI and the proposed algorithm respectively. The interpolation results of the proposed algorithm present the highest qualities.

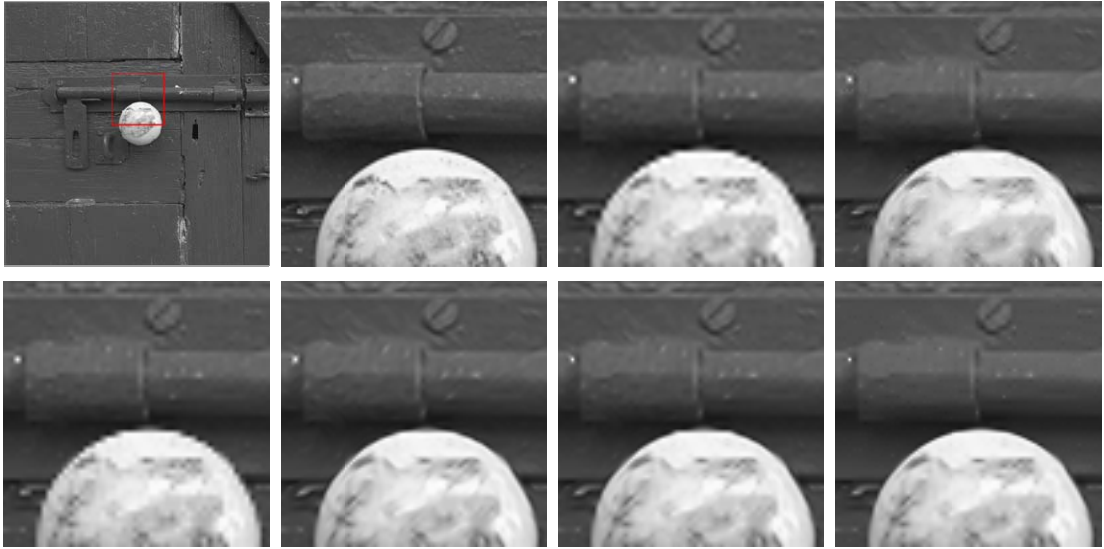

**Fig. 5**. The visual quality comparison on *Door*. Methods from left to right and then from top to bottom are original image, original block, and the blocks interpolated by Bicubic, NEDI, NLEDI, SAI, Robust SAI and the proposed algorithm respectively. The interpolation results of the proposed algorithm present the highest qualities.

## ACKNOWLEDGMENT

This work was supported in part by the Major State Basic Research Development Program of China (973 Program 2009CB320905), the Program for New Century Excellent Talents in University (NCET) of China (NCET-11-0797), and the National Science Foundation of China (NSFC) under grants 60803147 and the Fundamental Research Funds for the Central Universities (Grant No. HIT. BRETIII. 201221).